\documentclass[12pt]{article}

\usepackage{filecontents} %DT references in tex
\usepackage{times}
\usepackage{tabularx} %DT smart table
\usepackage{array} %included DT
\usepackage{subcaption} %included DT

\usepackage{caption} %DT
\usepackage[labelfont=bf]{caption} %DT
\usepackage{multirow} %DT
\usepackage{graphicx} %included DT

\usepackage[hidelinks]{hyperref}
\PassOptionsToPackage{hyphens}{url}\usepackage{hyperref}

\usepackage[]{url}
\usepackage{booktabs}
\usepackage{amsmath}

\author{Tanushkina Daria, Shevchenko Valeriy, Lukashevich Aleksander, \\ Bulkin Aleksandr, Grinis Roland, Kovalev Kirill, \\ Narozhnaia Veronika, Sotiriadi Nazar, Krenke  Alexander, Maximov Yury}

\title{Climate Change and Future Food Security: Predicting the Extent of Cropland Gain or Degradation}

\date{\today}

\usepackage[sorting = none]{biblatex}
\begin{document}

\baselineskip24pt

\maketitle

\begin{abstract}
Agriculture is crucial in sustaining human life and civilization that relies heavily on natural resources. This industry faces new challenges, such as climate change, a growing global population, and new models for managing food security and water resources. Through a machine learning framework, we estimate the future productivity of croplands based on CMIP5 climate projections on moderate carbon emission scenario. We demonstrate that Vietnam and Thailand are at risk with a 10\% and 14\% drop in rice production, respectively, whereas the Philippines is expected to increase its output by 11\% by 2026 compared with 2018. We urge proactive international collaboration between regions facing crop land gain and degradation to mitigate the climate change and population growth impacts reducing our society's vulnerability. Our study provides critical information on the effects of climate change and human activities on land productivity and uses that may assist such collaboration.

\end{abstract}

\section*{Introduction}

Climate change has a substantial impact on crop production, which poses risks to food security globally~\cite{cafiero2018food}. The Secretary-General of the United Nations has highlighted that Least Developed Countries are particularly vulnerable to these risks, especially given rising food and energy costs~\cite{Guterres_LDC2023}. While advances in technology have been made since the Industrial and Green Revolutions, climate change and weather variability remain the primary factors affecting crop production~\cite{jagermeyr2021climate}. Anthropogenic climate change exacerbates temperature and precipitation extremes, further compounding the issue \cite{zhou2023global}. Agricultural investments are highly challenging due to various financial and natural risks \cite{Food_security,shevchenko2023climate}, including nutrient price volatility, production losses, market fluctuations, political instability, regulatory changes, and supply chain disruptions. Althought observing land use changes can help to estimate how climate affects crops, allowing for more accurate risk evaluation and ultimately contributing to global food security. 

The above challenges are most crucial for countries with a high contribution of agriculture to gross domestic product (GDP) and employment rate. To help governmental agencies and policymakers address these challenges, we propose a high-fidelity data-driven approach to identify historical correlations between climate and arable land. Based on the widely recognized Coupled Model Intercomparison Project (CMIP) \cite{CMIP5}, we accurately forecast land utilization patterns and evaluate future production changes. This paper investigates countries with different levels of dependency on agriculture, specifically focusing on Central, South, and South-Eastern Asia, which present various levels of social and economic development, as outlined in Table~\ref{tab:agro_stat}. 

\begin{table}%[width=.9\linewidth,cols=3, pos=h]
    \caption{\textbf{Agricultural statistics in 2019 \cite{WB:2019_1,WB:2019_2}.}}
    \centering
    \begin{tabular}{lcc}%p{35mm} p{50mm} p{58mm}}
    \toprule
       Country  & \shortstack{Employment in\\agriculture (\% of total)} & \shortstack{Agriculture, forestry, and\\fishing, value added (\% of GDP)} \\ 
    \midrule
       Japan  & 3.4  & 1.0  \\
       Kazakhstan   & 14.9  & 4.5  \\
       Myanmar   & 48.9  & 21.4 \\
       Philippines      & 22.9    & 8.8 \\
       South Korea   & 5.1   &  1.7 \\
       Thailand & 31.4  & 8.1 \\
       Turkey & 18.1   & 6.4\\
       Uzbekistan   & 25.7  & 24.6  \\  
       Vietnam & 37.2 & 11.8 \\
    \bottomrule
    \end{tabular}
    \label{tab:agro_stat}
\end{table}

While many researches have attempted to establish a direct relationship between climate change and crop yields, ignoring the soil properties \cite{Sinnarong2022thailand, four_crops}, it is essential to note that previous imbalances in nutrient supply can seriously impact the effectiveness of mineral fertilizers \cite{Food_security}. Thus, our approach considers both climate and soil fertilizers usage to estimate the change in rice production. 

Despite notable progress in yield and crop modeling, many countries in the Asian region are understudied (except for India and China). Therefore, this gap presents an opportunity to produce regionally sensitive models that can account for region-specific factors and provide better quality results within areas of interest. Our study aims to promote investigations and modeling on a local scale using climate models, digital elevation models and tools like CMIP5 and Food and Agriculture Organization Corporate Statistical Database (FAOSTAT) \cite{CMIP5,FAOSTAT_prod, FAOSTAT_fert} to understand the impact of climate change on the environment. The latter will contribute to global sustainable agriculture efforts and improve the industry level of development, ultimately improving food security worldwide. 

Global efforts should focus on international collaboration between regions facing varying degrees of land gain or degradation proactively. This will play a critical role in sustainable agriculture development, allowing regions to combine resources, knowledge, and expertise to address the challenges of climate change, food security and beyond.

\section*{Results and discussion}
\subsection*{The status of arable lands is determined by bioclimatic variables}

\begin{figure*}[ht]
    \centering
    \includegraphics[width=\textwidth]{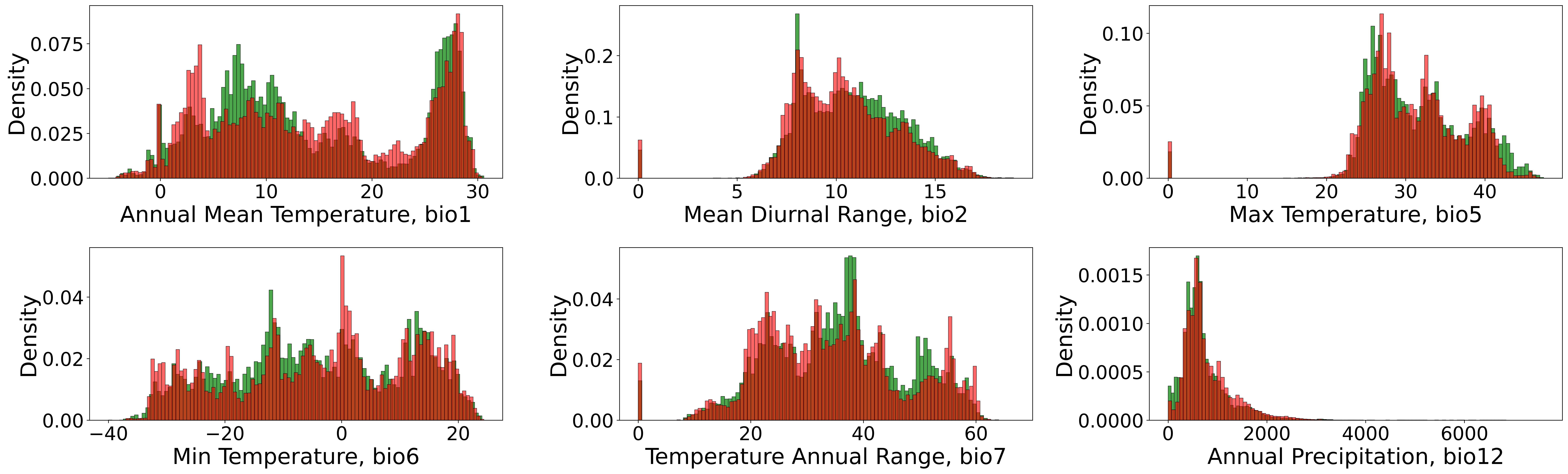}
    \caption{\textbf{Biovariables density distribution of samples lost (red) or acquired (green) cropland status.} Temperature is in $^{\circ}C$, precipitation is in mm.}
    \label{fig:lands_biovars_density}
\end{figure*}

The agricultural potential of croplands depends on various bioclimatic factors, such as temperature, rainfall, and other environmental conditions. To demonstrate this, we use the dataset prepared by Noce et. al. \cite{biovariables}. We will also use these data to predict the status of croplands. Fig. \ref{fig:lands_biovars_density} illustrates the density distribution of certain bioclimatic variables for the samples that either lost their crop production status (marked in red) or acquired it (marked in green) nine years later. The profiles of some features show a distinct shift, which is likely to aid in predicting the status of croplands. This shift indicates that relatively warm conditions, such as an average yearly temperature between $5^\circ$C and $10^\circ$C and a minimum temperature of the coldest month between $-20^\circ$C and $-10^\circ$C, are likely to result in the emergence of croplands. In contrast, harsher conditions can lead to their disappearance.

\subsection*{Effect of bioclimatic variables on croplands is delayed}

In agricultural settings, climatic variables have a substantial impact on the growth and development of crops. These effects may manifest immediately, such as when a hailstorm or flood damages crops, or they may be delayed, such as when soil loses vital nutrients due to prolonged changes in precipitation or droughts. 

Our study aims to predict arable lands' future status. We assume a land transformation happens a few years after actual climate conditions occur. In order to predict the status of land designated for cropland use, we determine an appropriate time delay in years between relevant features and the actual status. We then develope a machine learning tool that utilizes climate data to make these predictions (see below). Our evaluation of the tool's performance is dependent on the identified time delay value.

\begin{figure*}[ht]
    \centering
    \includegraphics[width=\textwidth]{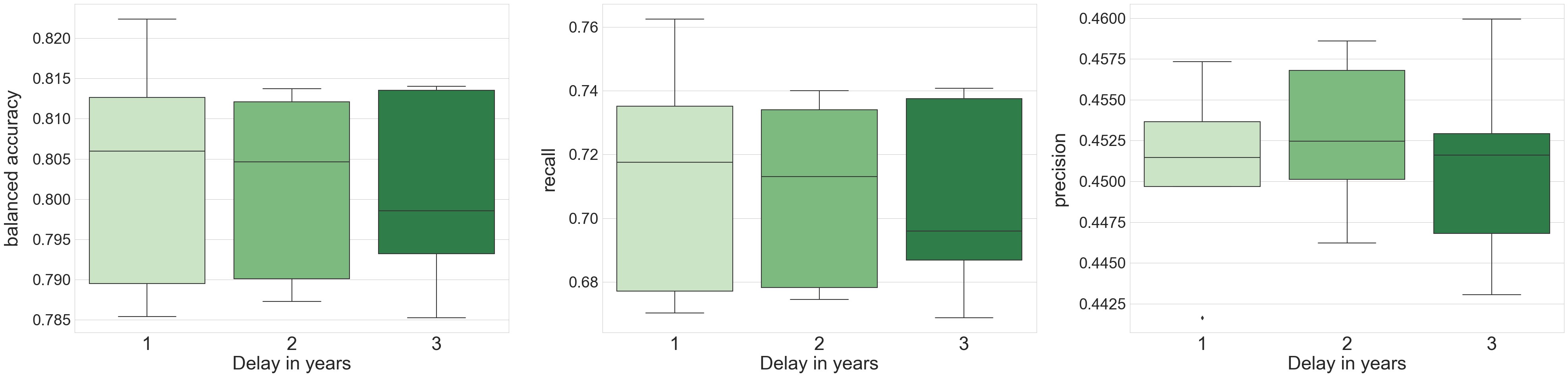}
    \caption{\textbf{Performance of the climate model with different land cover time delay.} N=6.}
    \label{fig:compare_lc_lag}
  \end{figure*}

The effectiveness of our tool is shown in Fig. \ref{fig:compare_lc_lag}, which evaluates the quality of cropland status prediction given a time lag. Classical performance metrics were used to assess the tool's performance using data bootstrapping, and the results suggest that a 1-year time delay is the most appropriate value for identifying the risks of cropland degradation. 

\subsection*{Agricultural use of fertilizers is expected to vary}

The significance of the delayed and immediate effects of climatic parameters on agricultural production also depends on other variables. Using fertilizers, a major attribute of the Green Revolution of the XX century, can help reduce the damage caused by weather and generally increases lands productivity. Based on agricultural statistics, one can identify trends in fertilizer consumption and predict future use.

\begin{figure}[ht]
    \begin{subfigure}{.32\textwidth} 
    \caption{}
    \includegraphics[width=\textwidth]{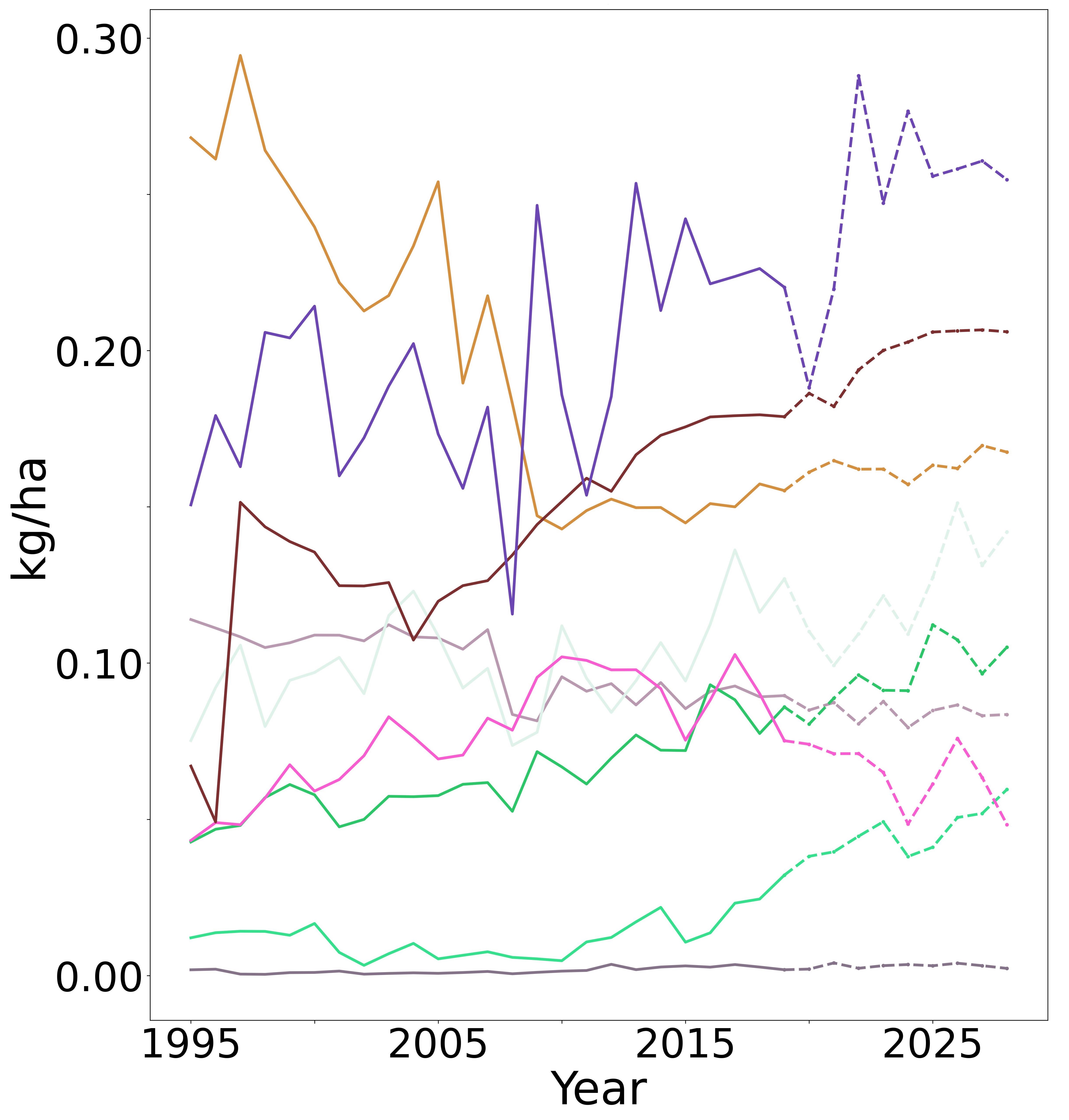}
\end{subfigure}   
\begin{subfigure}{.32\textwidth} 
    \caption{}
    \includegraphics[width=\textwidth]{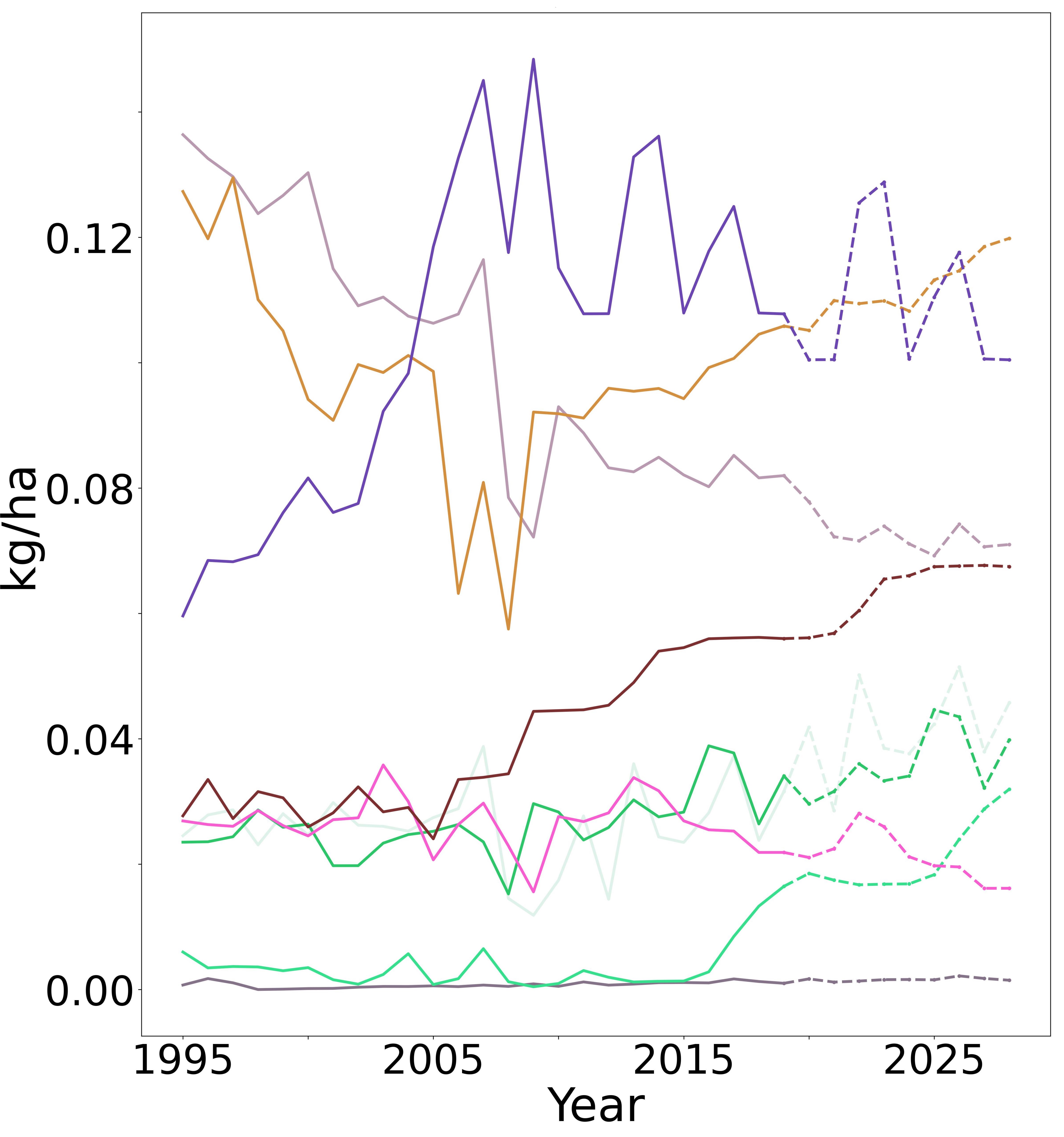}
\end{subfigure}
\begin{subfigure}{.32\textwidth} 
    \caption{}
    \includegraphics[width=\textwidth]{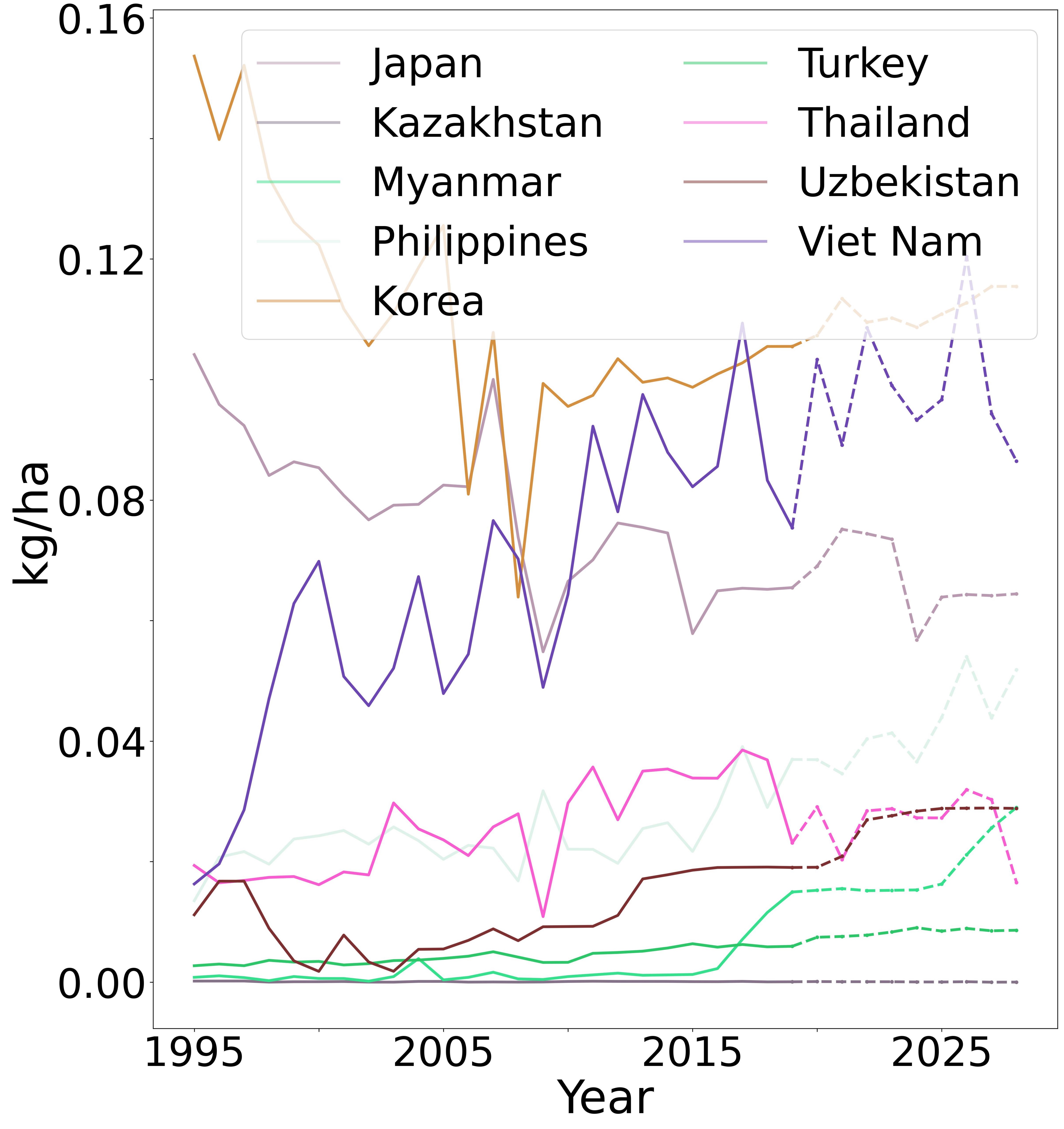}
\end{subfigure}  
    \caption{\textbf{Agricultural use of three primary feritilizer types. } 
The figure illustrates the historical fertilizers consumption before 2019 and the projected consumption from 2020 onwards, measured per unit of arable area. \textbf{a}, Nitrogen. \textbf{b}, Phosphorus. \textbf{c}, Potassium.}
    \label{fig:fert}
\end{figure}

Fig. \ref{fig:fert} demonstrates the agricultural consumption of three major types of fertilizers before 2019 and our forecasts (see Section \textbf{\nameref{sec:ricemodel}}) after 2020 using these data. Different countries show various dynamics of consumption for different types of fertilizers. These differences could potentially magnify the impacts of beneficial climate changes on certain regions, or conversely, exacerbate the ramifications of unfavorable changes. Under the neutral climatic change, the utilization of fertilizers alone may increase the productivity of croplands.

\subsection*{Accounting for \textit{unknowns}}

Artificial and natural complex systems, including agricultural systems, are subject to various unpredictable factors. While some of these factors may be beyond the domain of consideration, such as the possibility of a supernova explosion or the massive use of defoliants in a conflict, there are also domain-specific variables that are only implicitly present in climatic and agricultural statistics.

In actual agricultural practice, lands suitable for crop cultivation are selected based on their potential, either as a conscious decision of farmers who consider parameters not included in current statistics or analogously to the natural selection of the most suitable combinations of lands and other parameters. Therefore, considering past agricultural land use may help in the effort to predict the future status of croplands.

In order to confirm the validity of this approach, we constructed a variety of machine learning models utilizing classical and advanced techniques \cite{Kiwelekar2020dl_geo, gunen2021ml_wetlands}, incorporating climate and elevation data. Of these models, the XGBClassifier \cite{xgboost} outperforms its counterparts (see Table \ref{tab:metrics_land}). Subsequently, we proceeded by training the XGBClassifier model on data also encompassing land use dedicated to agricultural purposes for a period of five years prior to the prediction date. Hereafter, we will refer to this model as the ``model with memory''.

\begin{figure}
    \begin{subfigure}{0.5\textwidth} 
    \caption{}
    \includegraphics[width=\textwidth]{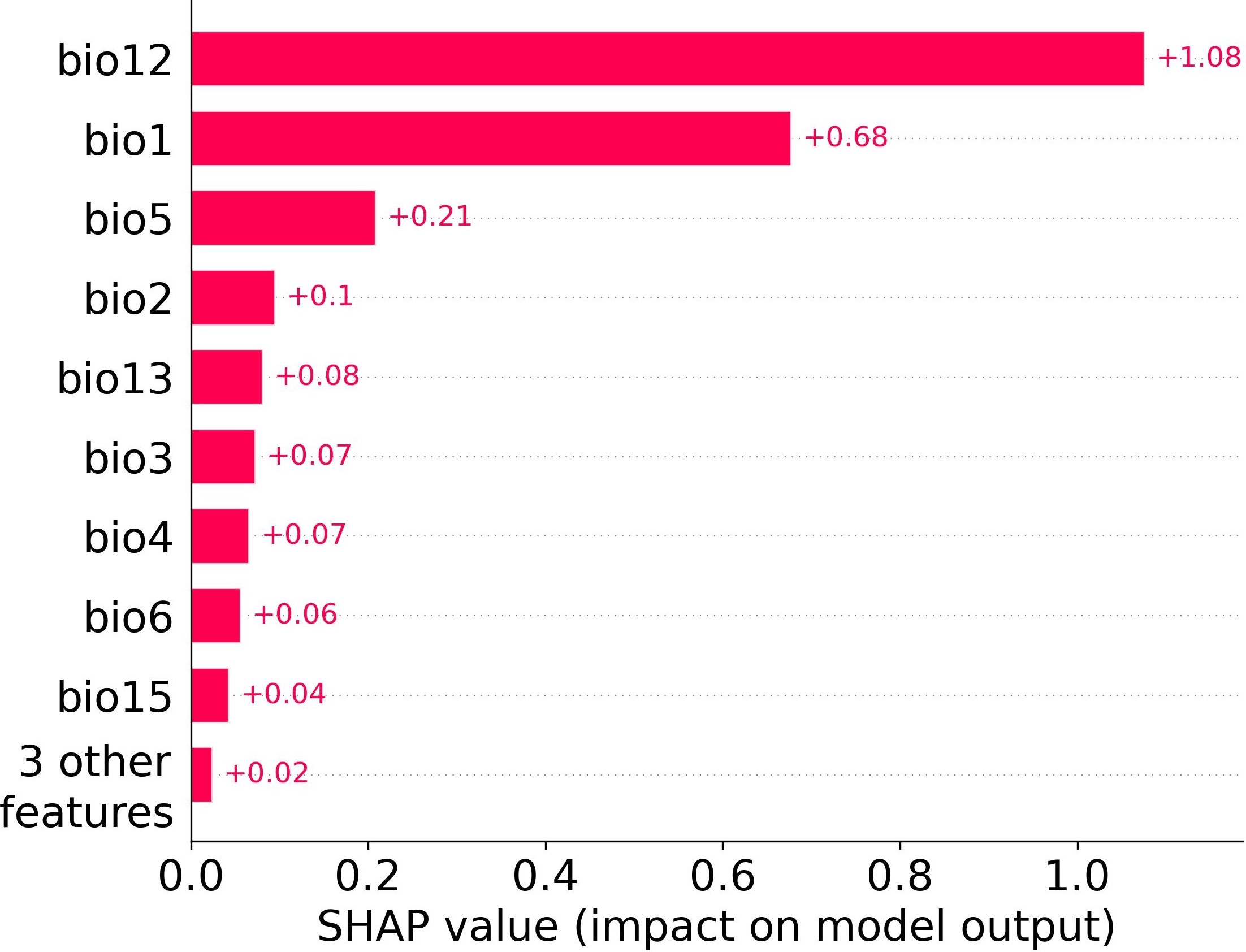}
    \end{subfigure}   
    \begin{subfigure}{0.5\textwidth}
    \caption{}
    \includegraphics[width=\textwidth]{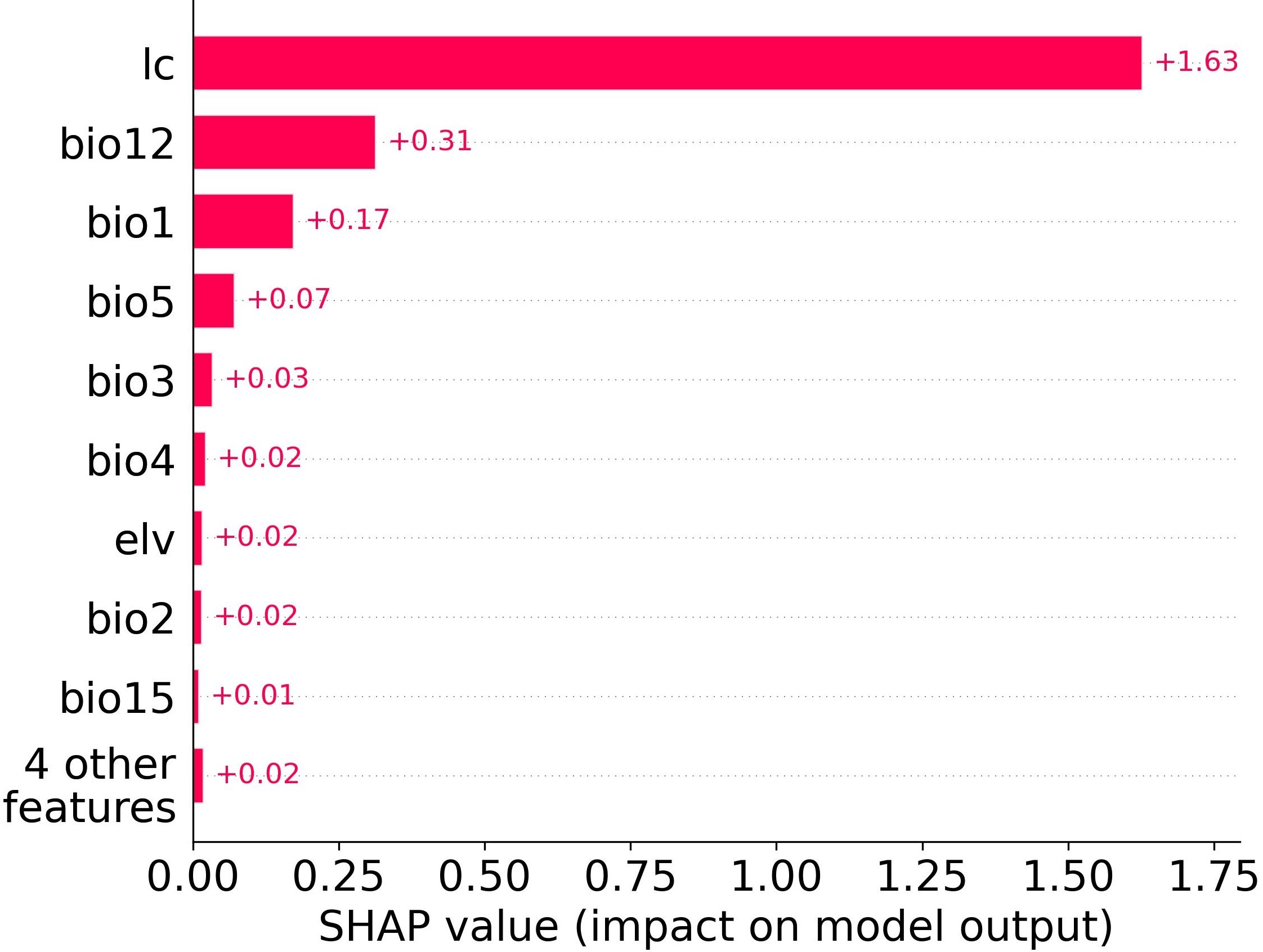}
    \end{subfigure}    
    \caption{\textbf{Feature importances evaluated with SHAP tool for the models with different set of features.} Features: lc -- land class 5 years prior; bio1, ... , bio12 -- see notations in Table \ref{tab:biovariables} \textbf{a}, Climate model. \textbf{b}, Climate model with memory.}
    \label{fig:models_feature_imp}
\end{figure}

In both cases, we set the time delay parameter $d=1$ for better model performance (see metrics in Table \ref{tab:metrics_land} and criteria in Section \textbf{\nameref{sec:modparam}}). 
The model with memory demonstrates superior results according to all the metrics we used. The feature importance analysis, performed with SHAP tool, using Shapley values to explain the output of the maching learning model with game theory \cite{Shap}, confirms that prior land use is the major factor contributing to the superior performance of the model with memory (Fig. \ref{fig:models_feature_imp}).

These findings highlight the importance of considering past agricultural land use when predicting the future status of croplands. This observation can be interpreted differently from various not mutually exclusive perspectives. Firstly, landowners may rely heavily on traditional farming methods, instead of adapting new techniques that account for shifts in climate conditions. Secondly, landowners may factor in local parameters that are absent from our climatic and agricultural dataset. Thirdly, landowners who utilize effective practices, regardless of a rational basis, can create a positive feedback loop through increased access to fertilizers or financial assets. 

\begin{table}[ht]
    \caption{\textbf{Classification metrics on test data.} Standard deviation was computed based on a sample size of N=10.}
    \centering
    
    \begin{tabular}{p{35mm} p{22mm} p{20mm} p{20mm} p{20mm} p{15mm}}
    \toprule
       
       Model & Classification threshold & Balanced accuracy & Precision & Recall  & ROC-AUC \\
    \midrule
      \multicolumn{6}{c}{\textit{Climate model}}\\
    \midrule
       Logistic Regression \cite{cox1958logreg} & 0.12 & 0.722 & 0.232 & 0.745 & 0.793 \\
       Random Forest Classifier \cite{ho1995randomforest} & 0.29 & 0.803 & 0.509 & 0.688 & 0.919 \\
       Naive Bayes \cite{zhang2004naiveb}& 0.40 & 0.757 & 0.265 & 0.777 & 0.821 \\
       MLP Classifier \cite{haykin1994mlp} & 0.24 & 0.797 & 0.455 & 0.695 & 0.909 \\
       AdaBoost Classifier \cite{schapire2013adaboost} & 0.50 & 0.786 & 0.487 & 0.656 & 0.907 \\
       CatBoost Classifier \cite{prokhorenkova2019catboost} & 0.25 & 0.813 & 0.554 & 0.694 & \bf{0.928} \\
       XGBClassifier \cite{xgboost}  & 0.27 & \bf{0.822} & 0.501 & \bf{0.734} & \bf{0.928} \\
       Convolutional Neural Network \cite{cun2015CNN}& 0.30 & 0.792 & \bf{0.579} & 0.509 & 0.812 \\%edit
    \midrule
       \multicolumn{6}{c}{\textit{Climate model with memory}}\\
      % \cmidrule{2-6}
    \midrule      
       \multirow{2}{*}{XGBClassifier} &
       \multirow{2}{*}{\shortstack{0.66\\        $\pm1.7{\times}10^{-2}$}} &
       \multirow{2}{*}{\shortstack{0.969\\        $\pm1.2{\times}10^{-6}$}} &
       % \tenpow[1.2]{-6}
       \multirow{2}{*}{\shortstack{0.945\\
       $\pm3.0{\times}10^{-6}$}} &
       \multirow{2}{*}{\shortstack{0.945\\
       $\pm2.8{\times}10^{-6}$}} &
       \multirow{2}{*}{\shortstack{0.990\\
       $\pm5.4{\times}10^{-5}$}}\\
       [0.5cm]
    \bottomrule
    \end{tabular}
    
    \label{tab:metrics_land}
\end{table}

\begin{figure}[h!]
\begin{subfigure}{\textwidth} 
    \caption{}
    \includegraphics[width=\textwidth]{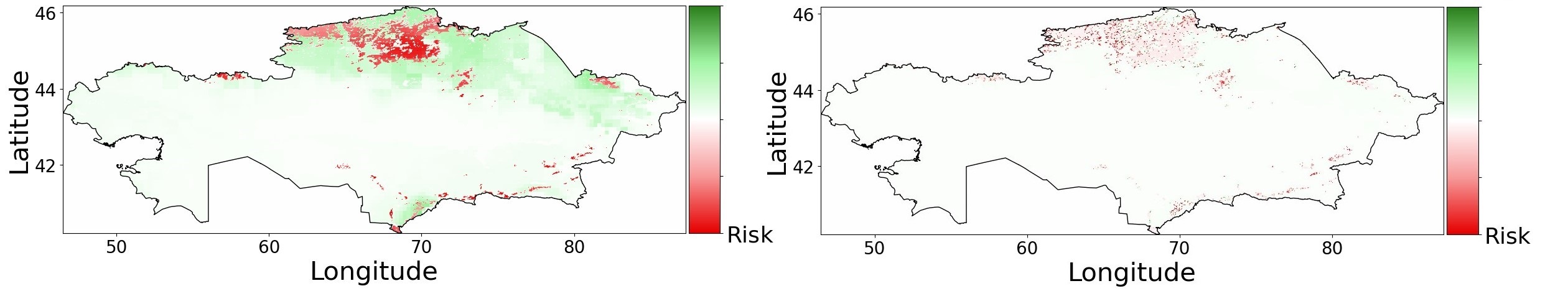}
\end{subfigure}   
\begin{subfigure}{\textwidth} 
    \caption{}
    \includegraphics[width=\textwidth]{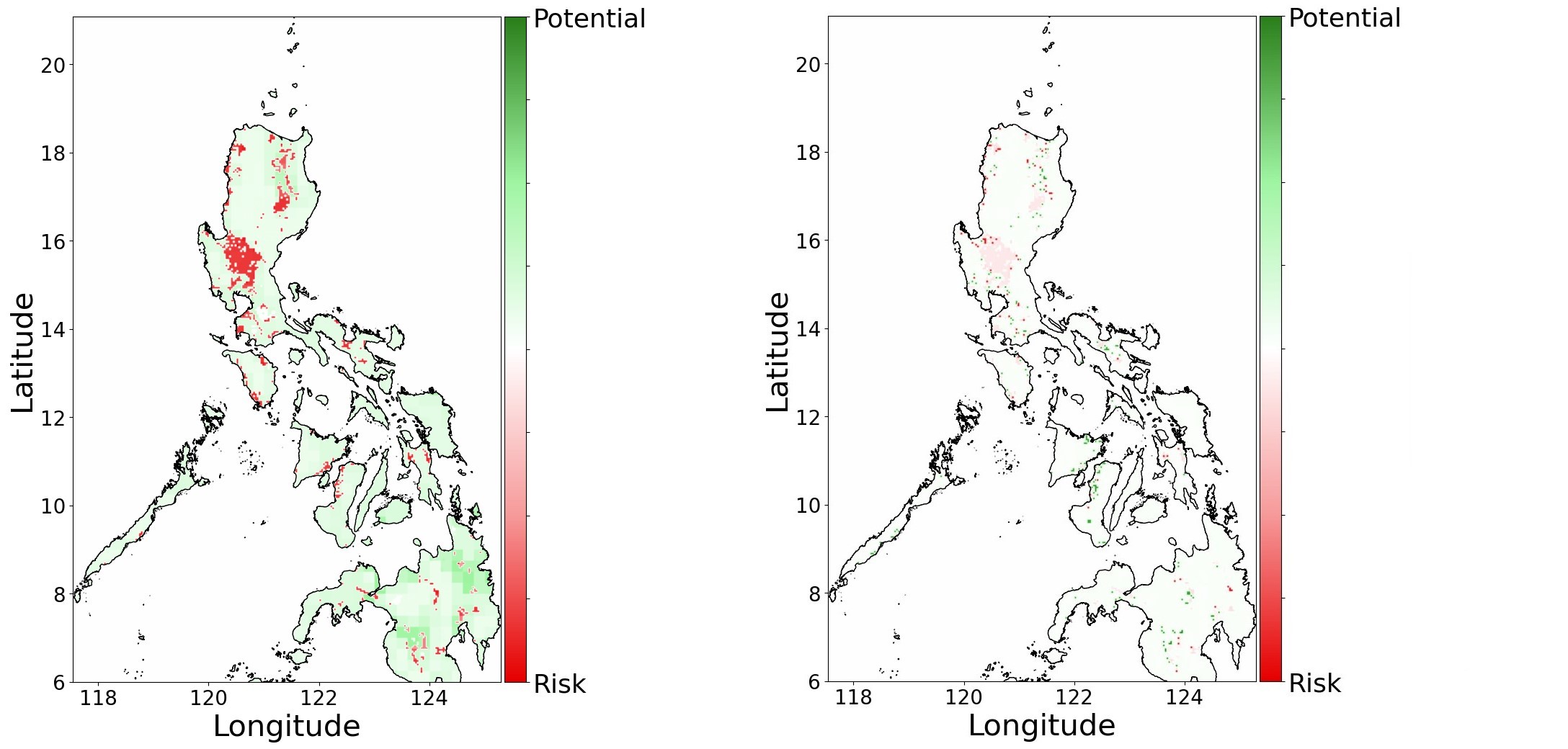}
\end{subfigure}
\begin{subfigure}{\textwidth} 
    \caption{}
    \includegraphics[width=\textwidth]{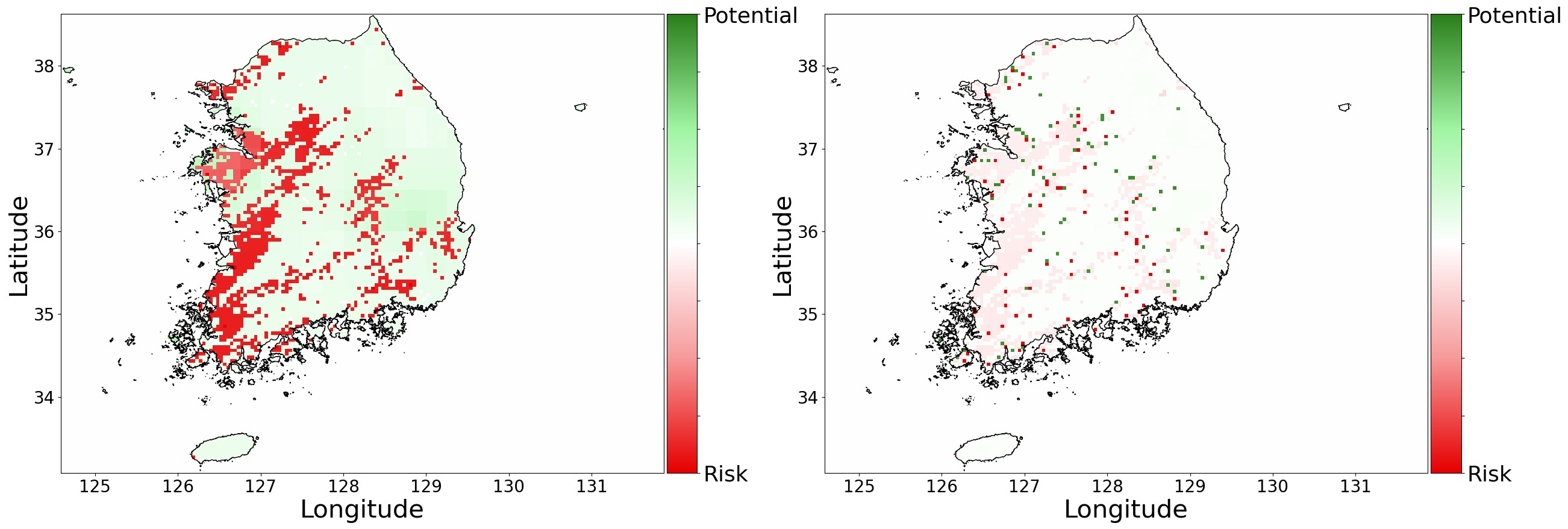}
\end{subfigure}  
    
\caption{\textbf{Expected changes in croplands in 2026 comparing with the historical year 2018 according to  climate model without (Left) and with memory (Right).} \textbf{a},   Kazakhstan. \textbf{b}, Philippines. \textbf{c}, Korea.}
\label{fig:changes}
\end{figure}

Our attempt to account for unknown factors in crop lands' status using the recorded land use history demonstrates that the model without memory generally overestimates changes in future crop lands' status. Fig. \ref{fig:changes} shows the potential switch in crop lands' status for three selected countries with diverse climates: Kazakhstan, Philippines, and South Korea, estimated using both models. The green color highlights the pixels with good potential, while the red color indicates those in danger for cultivation. The comparison of the two models allows us to identify the areas where present-day agricultural practices are no longer advantageous. Conversely, it also reveals areas where crop production is favored by climate change, and cultivation can be initiated with positive environmental outcomes.

Hereafter, we rely on a climate model with memory, which gives a moderate forecast. Since it uses the land status five years prior to the considered time, any prediction will be made in a short time horizon.

\subsubsection*{Yield changes due to climate change, fertilizers use, and other factors}

\begin{figure}
    \centering
    \includegraphics[width=\textwidth]{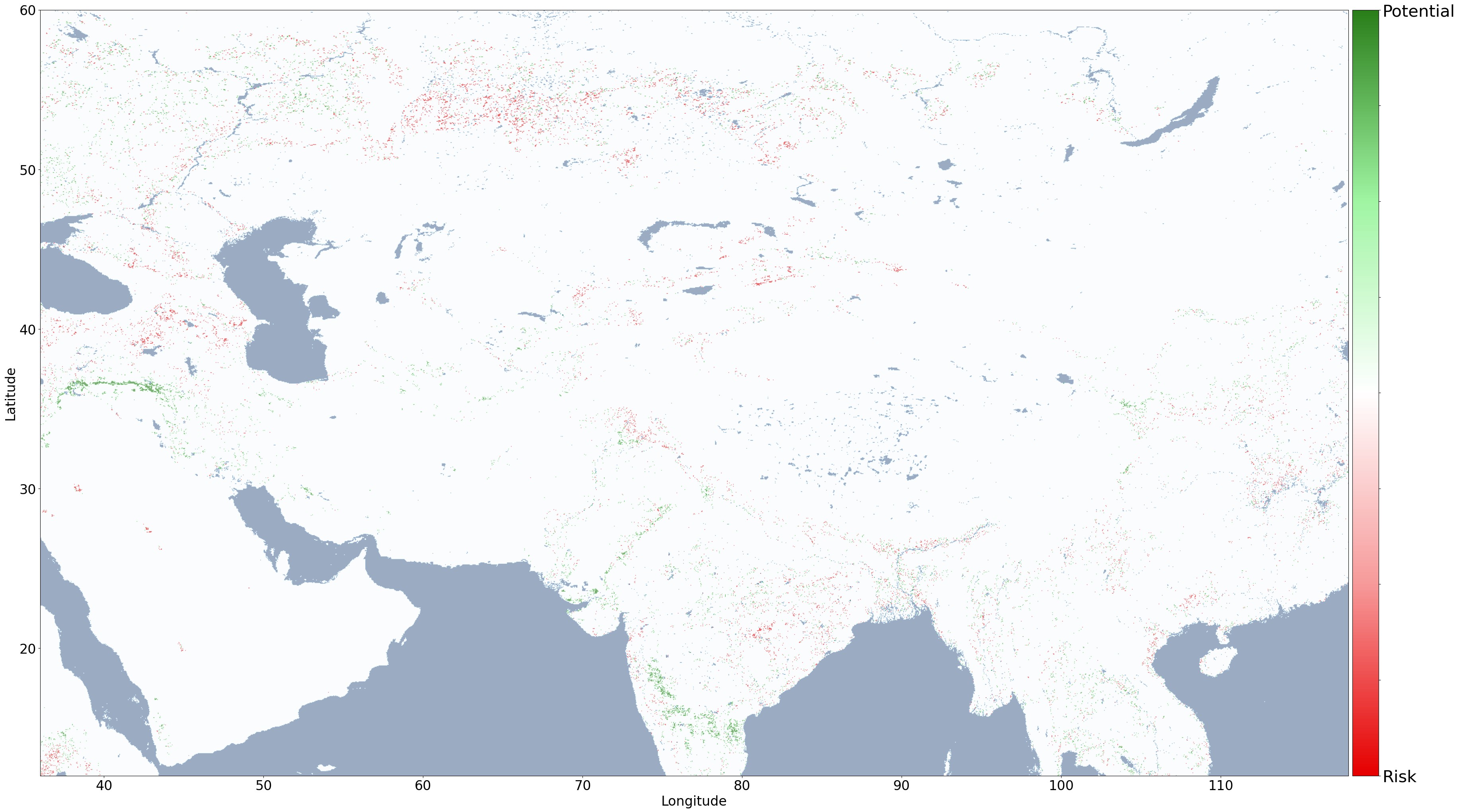}
    \caption{\textbf{Expected risky and potentially successful arable lands in Asia in year 2026 made with climate model with memory.}}
    \label{fig:world}
  \end{figure}

Whereas the creation of additional agricultural lands as a result of positive change is unlikely due to the conservative nature of the sector, any hostile conditions will inevitably result in reduced yields within affected regions. We examined the cultivation of rice to evaluate this approach. Fig. \ref{fig:world} shows the map with marked pixels having a high probability of crop status changes in the year 2026 compared to the year 2018 according to the climate model with memory. Figure \textbf{S1} contains high resolution picture. Green color highlights the pixels with good potential, and red color -- those with danger for cultivation. Table \ref{tab:changes} demonstrates the results of rice yield modeling utilizing the climate model with memory explained in Section \textbf{\nameref{sec:ricemodel}}. 

A negative yield or production value indicates a decrease, whereas a positive value represents an increase. Our yield model shows $R^2 = 0.944$  and mean absolute percentage error $MAPE=5.7\%$ on test data. It is important to note that these findings adopt a cautious approach and only account for risky lands, thereby disregarding prosperous regions. This study focuses on identifying potential risks rather than proposing development strategies. According to our findings, Thailand and Vietnam face severe threats in rice production, while the Philippines is expected to experience growth. Identifying region-specific factors contributing to these trends is challenging, as climate conditions and soil fertilizer levels vary independently. Thus, the efficacy of fertilizers may vary, and their impact can range from negligible to significant.

\begin{table}%[ht]
\caption{\textbf{Expected percentage change of yield and rice production according to the climate model with memory in 2026 compared with the historical year 2018. }}
\centering

    \begin{tabular}[width=\textwidth]{p{21mm} wc{18mm} wc{18mm} wc{18mm}| wc{18mm}wc{18mm}}
   \toprule
        \multirow{2}{*}{Country}&\multirow{2}{*}{\shortstack{Yield\\change,\%}}&\multicolumn{2}{c}{General analysis}&\multicolumn{2}{c}{Analysis with rice mask}\\

    \cmidrule{3-6}
      &&\shortstack{Lands at \\risk, \%} &
      % \shortstack{Model un-\\certainty, \%}&
      \shortstack{Production\\change, \%}&{\shortstack{Lands at \\risk, \%}} & \shortstack{Production\\change, \%} \\
   \midrule
       Japan  & -2.1  & 10.4& -12.3&-&-\\
       Kazakhstan   & -4.7  & 16.4 & -20.4&-&-\\
       Myanmar   & -0.8  & 3.2 & -4.0&0.9&-1.7\\
       Philippines      & 12.8  & 7.4& 4.4&1.8&10.8\\
       South Korea   & -4.7  & 6.5& -10.9&4.1&-8.7\\
       Thailand & -13.8  & 1.9   & -15.4&0.4&-14.1\\
       Turkey & -9.2  & 9.7& -18.1&-&-\\
       Uzbekistan   & 9.8  & 2.4 & 7.2&-&-\\  
       Vietnam   & -6.0  & 11.9  & -17.3&4.6&-10.4\\  
   \bottomrule
    \end{tabular}
    
    \label{tab:changes}
\end{table}

The primary constraint of this study pertains to the grid roughness. The spatial resolution employed (roughly 4500 m in cell length) is larger than the field size, resulting in several diverse areas within the same pixel. Additionally, this grid is uniform and does not correspond to the actual shapes of the fields. 

Lastly, our modeling relies on the datasets listed in Table \ref{tab:datasets}. Some of these are the results of modeling studies, which inherently approximate natural phenomena and, therefore, are imprecise. It influences our model and should be considered when interpreting the model outcomes.

However, there is room for improvement. Various studies conducted under the CMIP5/ CMIP6 project can assist in overcoming the limitations of mathematical simulation in reproducing natural processes. Global-scale processes are incredibly complex. Accurate reproduction of such processes with mathematical simulations is still impossible. Each model has advantages and disadvantages in replicating changes occurring on land, in the atmosphere, in permafrost, or above the ocean. The appropriate work direction could be collecting the region-specific CMIP models of reasonable quality into an ensemble \cite{CMIP_ensembles}. 
This approach may separate the strengths and weaknesses of the models to increase the reliability of the outcomes.

\section*{Materials and methods}
\label{sec:methods}

\subsection*{Data and preprocessing}
\label{sec:data}
In this study we develop a model employing several open datasets detailed in Table \ref{tab:datasets}. We get remote sensing data with Google Earth Engine \cite{GoogleEE}, and assume elevation to be invariant through all the considered time. The land classification is based on the University of Maryland classification \cite{Land_cover,hansen2000global}. We transform the land cover to binary classification with crops (labeled as class 12 in the source) and non-crops (all other classes).

\begin{table}[h!]
    \caption{\textbf{The datasets desctiption.}}
    \centering
    \begin{tabular}[width=\textwidth]{p{4.2cm} p{4.3cm} p{1.8cm} p{1.8cm} p{2.0cm}}
       \toprule
    Dataset Name & Variable & Time coverage & Spatial resolution & Temporal resolution \\
    \midrule

    NASA SRTM Digital Elevation \cite{elevation}  &  Elevation & - & 1 arcsec & - \\[1cm]

    MCD12Q1 Land Cover Type\cite{Land_cover} & LC \underline{ }Type2 & 2001–2020 & $\frac{1}{20}^\circ$ \newline & Yearly\\[1cm]
      
    TerraClimate \cite{TerraClimate} & Minimum and maximum temperatures, precipitation & 1958–2021 & $\frac{1}{24}^\circ$ &  Monthly \\[1.3cm]

    CMIP5 \cite{CMIP5} & 
    Monthly mean of the daily-minimum and daily-maximum near-surface air temperatures, sum of precipitation at surface & 1950–2100 &  $\frac{1}{2}^\circ-\frac{3}{2}^\circ$ &  Daily \\[3cm]
    % \vspace{20.5mm}
    
    NESEA-Rice10 \cite{Han2021riceMap}& 
    Paddy rice map& 2017–2019 &  $0.0001^\circ$&  Yearly \\[0.6cm]
    % \vspace{20.5mm}
    Global Administrative Areas \cite{GADM} & Administrative boundaries of the countries  & 2022 &-&-\\[1cm]
    % \vspace{0.5mm}
    Food and Agriculture Organization Corporate Statistical Database \cite{FAOSTAT_prod, FAOSTAT_fert} & Rice (production quantity), rice (area harvested), fertilizers by nutrient (agricultural use) & 1961–2021 & - & Yearly\\[1cm]
    \bottomrule
    \end{tabular}
    
    \label{tab:datasets}
\end{table}

Historical climate data come from TerraClimate source \cite{TerraClimate}. Its original spatial grid serves as the foundation to project all other data. %{\color{blue} I did not get it} {\color{teal} edited}
Using all the datasets in a single model requires aligning them in space and time.

We consider the future climate data from various CMIP5 simulations based on multiple evaluations conducted by different groups \cite{CMIP5_comparison,CMIP5_comparison_pr} to ensure the high-fidelity and robustness of the results. The careful selection is crucial to the outcome of the model. This study focuses on Central Asia, Tibetan, East Asia, South Asia, Southeast Asia, Siberian, and Mediterranean regions as they are labeled in different sources  \cite{Giorgi2002_regions}.  Table \ref{tab:CMIP5_models} lists  simulations under the moderate Representative Concentration Pathway (RCP) 4.5 scenario of greenhouse gas concentration trajectory employed in this study.

\begin{table}[h]
    \caption{\textbf{CMIP5 simulations used in this study.}}
    \centering
    \begin{tabular}{p{40mm} p{110mm}}
   \toprule
        Model name & Institution\\
   \midrule
        CNRM-CM5 & Centre National de Recherches Météorologiques, France \\
        GFDL-CM3 & NOAA Geophysical Fluid Dynamics Laboratory, United States \\
        MPI-ESM-MR & Max Planck Institute for Meteorology, Germany \\
   \bottomrule
    \end{tabular}
    
    \label{tab:CMIP5_models}
\end{table}

The climate projections we are examining are based on a daily temporal resolution. During the preprocessing stage, we calculate the mean maximum and minimum temperatures, as well as cumulative precipitation figures, for each month.  Furthermore, historical and future climate data are used to calculate a few bioclimatic variables, presented in Table \ref{tab:biovariables}, annually. These variables are informative in determining patterns in temperature and precipitation variations. They help to reduce observation noise, computational requirements and training bias. 

\begin{table}[ht]
    \caption{\textbf{Bioclimatic variables.}}
    \centering
    \begin{tabular}{p{25mm} p{70mm}}
   \toprule
        Biovariable & Description\\
   \midrule
        bio1 & Annual mean temperature \\
        bio2 & Mean diurnal range \\
        bio3 & Isothermality \\
        bio4 & Temperature seasonality \\
        bio5 & Max temperature of warmest month \\
        bio6 & Min temperature of coldest month \\
        bio7 & Temperature annual range \\
        bio12 & Annual precipitation \\
        bio13 & Precipitation of wettest month \\
        bio14 & Precipitation of driest month \\
        bio15 & Precipitation seasonality \\
   \bottomrule
    \end{tabular}
    
    \label{tab:biovariables}
\end{table}

Evaluation of the initial data revealed an imbalanced distribution of classes, with an average of 11\% of all lands assigned to crops and up to 3\% of them subject to changes in the next few years. The area of interest is considered as a uniform spatial grid, where each pixel has elevation value, bioclimatic values, and designated land class as the target label.

\subsection*{Model}\label{sec:algo}

Extreme Gradient Boosting Classifier XGBClassifier \cite{xgboost} was chosen as a machine learning baseline because its performs better than other tools when applied to the same data in our pilot study (Table \ref{tab:metrics_land}, also see \cite{lad2022factors}). 

At the first step, all the features are used for training with grid search  and StratifiedKFold cross-validation among several regularizations and decision tree parameters. The procedure of choosing optimal parameters is given in Section \textbf{\nameref{sec:modparam}}.

The proposed model involves collecting two matrices --- $X_{train}$ and $y_{train}$ --- from data in different years. The $y_{train}$ values are collected from data delayed by an integer parameter $d$ in time. For example, $X_i$ corresponds to a year $i$, while $y_i$ corresponds to the year $i + d$. We select the parameter $d$ based on the distribution of relevant metrics derived from data collected across multiple time periods (for details see Section \textbf{\nameref{sec:modparam}}). %{\color{blue} summarize the justification}.{\color{teal}done}
Using this collected data, we train a binary classifier to predict the probability of assigning either class 1 (arable land) or class 0 (not arable land) to a specific sample, which is described with features listed in Table \ref{tab:biovariables} along with elevation.

\section*{Numerical experiments}\label{sec:exper}

\subsection*{Data Analysis}

Our study focuses on the proposed approach and its application in Central, South, and South Eastern Asia. We cover a diverse range of countries with varying levels of social and economic development, including Japan, Kazakhstan, Myanmar, the Philippines, South Korea, Thailand, Turkey, Uzbekistan, and Vietnam. The region of the study is limited to a latitude range of $35^\circ$N to $60^\circ$N and a longitude range of $27^\circ$E to $166^\circ$E. The spatial resolution utilized is $1^\circ/24$ , which was determined through the algorithm described in Section~\textbf{\nameref{sec:data}}.

A preliminary data analysis showed that certain samples that had changed their labels from crop to non-crop and vice versa over the years modeled. Fig. \ref{fig:lands_biovars_density} shows the density distribution of particular biovariables for the samples that lost their crop production status (marked in red) or acquired it (marked in green) nine years later. The figure depicts a clear shift in the profiles of some features, which is likely to assist the model to distinguish between sample groups. 

\subsection*{Model Parameters}\label{sec:modparam}

We performed grid search in order to estimate optimal parameters for further modeling. Table \ref{tab:gridSearch} displays the initial parameter sets and the optimal values that we chose.

\begin{table}[h]
    \caption{\textbf{Grid search parameters.}}
    \centering
    \begin{tabular}{p{30mm} cc}%p{20mm} p{40mm}}
   \toprule
        Parameter & Tested set & Chosen value\\
   \midrule
        $\textup{reg\_alpha}$ & $0$, $10^{-1}$, $1$ & $10^{-1}$\\
        $\textup{reg\_lambda}$ & $1$, $10$, $10^2$ & $1$\\
        $\textup{max\_depth}$& $3$, $4$, $5$ & $5$ \\
        $\textup{learning\_rate}$& - & $0.02$ \\
        $\textup{n\_estimators} $& - & $200$ \\
   \bottomrule
    \end{tabular}
    \label{tab:gridSearch}
\end{table} 

Fig.~\ref{fig:lands_biovars_density} shows the distribution of most essential features. Aside from climate data and land class, the models listed in Table~\ref{tab:models} include elevation (elv) and land class 5 years prior (lc). The inclusion of ``memory'' within the name indicates that historical land classes, i.e., prior land usage of this land, were also used as a part of its feature space.

\begin{table}[h]
    \caption{\textbf{Models with features included.} Biovariables are listed in Table \ref{tab:biovariables}.}
    \centering
    \begin{tabular}{p{50mm} p{40mm}}
    \toprule
        Model Name & Features\\
    \midrule
        Climate model &11 biovariables, elv \\
        Climate model with memory & 11 biovariables, elv, lc \\
    \bottomrule
    \end{tabular}
    
    \label{tab:models}
\end{table}

In Section \textbf{\nameref{sec:algo}}, we define the data collection process with a time delay of $d$ years to represent the number of years required for crop owners to exclude land from cultivation due to climate conditions. Fig. \ref{fig:compare_lc_lag} displays classification metrics for various $d$ values. 

The classification threshold serves as a decision threshold that maps the classifier output probability of a sample being assigned to class 1 (presence of crops) to its actual binary category. Balanced accuracy is a performance metric that measures the percentage of correct predictions with respect to the share of each class, making it particularly useful when dealing with imbalanced classes where one class is underrepresented compared to the other. In this study, we use balanced accuracy to evaluate the performance of our classifier in distinguishing between the presence and absence of crops (denoted as class 1 and class 0, respectively). We estimate the precision (the tendency not to predict false croplands) and recall (the ability not to predict false non-croplands) using the optimal threshold based on maximizing the $F$-Measure --- the harmonic mean of precision and recall. 

As our study primarily focuses on the potential decline in soil productivity, recall appears to be of greater significance. Thus, the results highlighted in Fig. \ref{fig:compare_lc_lag} lead us to choose parameter $d=1$. We acquire the training set using TerraClimate data from 2006 to 2008 and land classes from 2007 to 2009. To avoid any potential data leakage, we take great care in selecting the train and test data. Specifically, we ensure that the land class in any given year is never used as both a label for training and a feature for testing.

Fitted model uses CMIP5 climate projections to make a forecast. Phase 5 is chosen since it has a better correspondence in temperature with recently observed data \cite{carvalho2022well}. We assume that the suitability of climate models may vary depending on the chosen climate zone. To improve consistency, we create an ensemble prediction by averaging the probabilities of each output among the simulations listed in Table~\ref{tab:CMIP5_models}.

\subsection*{Rice yield model}\label{sec:ricemodel}

To assess the potential effects of crop yield degradation, the separate regression model estimates the connection between climate, consumption of fertilizers, time trends, and yield as the target variable. We develop this approach to capture the link between specific social, economic traits and climate conditions. Climate features of the yield model include values of minimum and maximum for temperatures and precipitations, calculated as monthly means as well as variances of these values in the monthly distribution. We use climate data collected from the TerraClimate source within national borders that were acquired from the Global Administrative Areas dataset (see Table \ref{tab:datasets}). This approach yields 72 climate features in total. 

Fertilizer data include the agricultural use of nitrogen $N$ (in various chemical forms), potash $K_2O$, and phosphate $P_2O_5$ per country from 1995 to 2019. Fig.~\ref{fig:fert} illustrates their agricultural use $F$ until 2019, as indicated in Table \ref{tab:datasets}. The forecast for the future year $y$ is generated separately for each country $c$  and for each fertilizer with a simple 10-year  shift of the previously observed curve into the future:
\begin{align}
\begin{split}
F_{cy} = F_{c(y-10)} + s,\\
s = F_{c2019} - F_{c2009}.
\end{split}
\end{align}

We utilize the specific year values $F$ for nitrogen $N$, potash $K_2O$, and phosphate $P_2O_5$ to serve as three features in the modeling of yield. Country-specific time trends are also factored into the model. Mathematically, we set the functional dependence and estimate the unknown coefficients as follows:
\begin{align}
\begin{split}
    Y_{cy} = {}& w_1pr_{cmy}+w_2pr^{var}_{cmy}+w_3tmax_{cmy}+ \\ 
    & +w_4tmax^{var}_{cmy}+w_5tmin_{cmy}+w_6tmin^{var}_{cmy}+ \\
    & + w_7F^N_{cy} + w_8F^{P_2O_5}_{cy}+ w_9F^{K_2O}_{cy}+ \\
    & + \psi_{c}(y - 1995) + \epsilon_c + \gamma,
\end{split}
\end{align} 
where
\begin{itemize}
\item $Y$ is rice yield,
\item $c$, $m$, $y$ represent country, month and year respectively,
    \item $pr$ and $pr^{var}$ are precipitation level and its variance,
    \item $t_{\max}$ and $t_{\max}^{var}$ are maximum temperature and its variance,
    \item $t_{\min}$ and $t_{\min}^{var}$are minimum temperature and its variance,
    \item $F$ are fertilizer consumptions,
    \item $\psi$ are time trends,
    \item $\epsilon$ and $\gamma$ are the error terms.
\end{itemize}

We train the model with the annual data spanning from 1996 to 2017, and test its performance for the subsequent years of 2018 to 2019. The collected data for these years boast complete coverage within our area of interest, allowing for a comprehensive analysis.

To determine the yield as a target variable, we divide the rice production of a specific country by its corresponding cultivation area, with both values sourced from the FAOSTAT data (see Table \ref{tab:datasets}). National statistics and climate data are utilized to obtain the necessary information for calculating yield forecasts. We then apply this regression model to estimate future rice yields in a given country. When combined with the expected reduction in area, it effectively predicts the rice production.

Available paddy rice dataset \cite{Han2021riceMap} offers an opportunity to improve the precision of rice land assessment in Myanmar, Philippines, South Korea, Thailand, and Vietnam. When using it as a mask for crop fields, we refer to this as the ``rice mask'' analysis and demonstrate the significance of utilizing these data in current research. By employing this mask, we enhance the accuracy of climate change impact estimation on rice production, compared to the general analysis that does not consider the specific location of rice fields.

\section*{Conclusion}\label{sec:conclusion}

This work presents evidence for the impact of climate on croplands. The study utilized a basic climate model that gathered biovariables based on historical climate data. These biovariables, such as annual mean temperature, maximum temperature, and annual precipitation, were used to predict the presence or absence of cropland in the future based on climate projections. The results showed that even moderate modeling suggests a high likelihood of severe conditions for growing crops in Thailand, and Vietnam. Consequently, these lands will either undergo a land transformation or experience a notable drop in yield.

Furthermore, the study allows for comparing neighboring regions. Underrated clusters were identified where crop potential is high, but the share of cultivated fields is low. This finding calls for local policy changes and investor initiatives, which could be used for regional development planning, creating agricultural road maps, water management, and more.

In addition to business motivations, the topic has a more comprehensive scope as it relates to global food security. Climate change is responsible for rearranging conventional food supply chains on regional and international scales. Predictions based on the findings of this study can help take measures to mitigate the impact of climate change on food security before actual transformations occur.

We also believe that this study complements various studies on the impact of natural disasters such as floods or hail to food security~\cite{morozov2023cmip,abramov2023advancing,mozikov2023accessing} often combines CMIP models with the analysis of extremes~\cite{morozov2023cmip}.

% \section*{{\color{red}Acknowledgments}}
% We acknowledge the World Climate Research Programme’s Working Group on Coupled Modelling, which is responsible for CMIP, and we thank the climate modeling groups (listed in table \ref{tab:CMIP5_models} of this paper) for producing and making available their model output.
% \section*{{\color{red}Fundings}}
% Include acknowledgments of funding, any patents pending, where raw data for the paper are deposited, etc.

% \section*{Author Contributions}
% Conceptualization: AL, DT, AK, YM\\
% Methodology: AL, AK, NL, AB, VS\\
% Investigation: DT, VS, AL, NL\\
% Visualization: DT\\
% Supervision: IB, NS\\
% Writing—original draft: DT, IB, AL, VS\\
% Writing—review and editing: IB, DT

\section*{Competing Interests}
The authors declare that they have no competing interests.

\section*{Code availability}
All data references and source code needed to evaluate the conclusions in the paper are publicly available through Zenodo at \url{https://doi.org/10.5281/zenodo.8033309}.

%\section*{Supplementary materials}
% Materials and Methods\\
% Supplementary Text\\
%Fig. S1\\
% Tables S1 to S4\\
% References \textit{(4-10)}

\printbibliography
\end{document}